\documentclass{IEEEoj}
\usepackage{cite}
\usepackage{amsmath,amssymb,amsfonts}
\usepackage{algorithmic}
\usepackage{graphicx,color}
\usepackage{textcomp}
\usepackage{subfigure}
\usepackage{threeparttable}
\usepackage{multirow}
\usepackage{booktabs}
\usepackage{array}
\usepackage{ragged2e}
\usepackage[T1]{fontenc}
\usepackage{xurl}
\usepackage[hidelinks]{hyperref}
\def\BibTeX{{\rm B\kern-.05em{\sc i\kern-.025em b}\kern-.08em
    T\kern-.1667em\lower.7ex\hbox{E}\kern-.125emX}}
\makeatletter
\def\ps@headings{\def\@oddhead{\vbox{\vspace{17pt}\hsize\textwidth\hbox{\rfxfont\rightmark\hfill}\hfill\par
\smallskip\noindent\hbox to \textwidth{\vrule width\textwidth height.3pt depth0pt}}}%
\def\@evenhead{\vbox{\vspace{17pt}\hsize\textwidth\hfill\hbox{\hfill\rhfont\leftmark}\par
\smallskip\noindent\hbox to \textwidth{\vrule width\textwidth height.3pt depth0pt}}}%
\def\@oddfoot{\hfill\rffont\thepage}\def\@evenfoot{\rffont\thepage\hfill}}
\def\ps@plain{\def\@oddhead{\vbox{\vspace{17pt}\hsize\textwidth\hbox{\rhfont\leftmark\hfill}\hfill\par
\smallskip\noindent\hbox to \textwidth{\vrule width\textwidth height.3pt depth0pt}}}%
\def\@evenhead{\vbox{\vspace{17pt}\hsize\textwidth\hfill\hbox{\hfill\rhfont\leftmark}\par
\smallskip\noindent\hbox to \textwidth{\vrule width\textwidth height.3pt depth0pt}}}%
\def\@oddfoot{\hfill\rffont\thepage}\def\@evenfoot{\rffont\thepage\hfill}}
\makeatother
\AtBeginDocument{\definecolor{ojcolor}{cmyk}{1,0.45,0,0.18}%
\definecolor{ojcolor2}{cmyk}{0,0.91,0.81,0.19}%
\pagestyle{headings}\thispagestyle{plain}}
\begin{document}

\receiveddate{XX Month, XXXX}
\reviseddate{XX Month, XXXX}

\title{\textcolor{ojcolor2}{DTCO of NOR-Type IGZO FeFETs for 3D Heterogeneous AI Memories: A Read-Centric Perspective}}

\author{YANG XIANG$^{1}$, ZHUO CHEN$^{1}$, NICOL{\`{O}} RONCHI$^{1}$, ARVIND SHARMA$^{1}$, 

FERNANDO GARC\'{I}A-REDONDO$^{2}$, SUBHALI SUBHECHHA$^{1}$, ATTILIO BELMONTE$^{1}$, 

MAARTEN ROSMEULEN$^{1,3}$, GOURI SANKAR KAR$^{1}$, DWAIPAYAN BISWAS$^{1}$, 

JAN VAN HOUDT$^{1,4}$}

\affil{imec, Leuven, 3001 Belgium}
\affil{imec, Cambridge, CB1 2JD U.K.}
\affil{Department of Electrical Engineering, KU Leuven, Leuven, 3001 Belgium}
\affil{Department of Physics and Astronomy, KU Leuven, Leuven, 3001 Belgium}

\corresp{CORRESPONDING AUTHOR: Yang Xiang (e-mail: Yang.Xiang@imec.be)}
\authornote{This work was supported by Imec's Industry Affiliation Program (IIAP).
\textit{The paper has been submitted to the IEEE for possible publication. Copyright may be transferred without notice, after which this version may no longer be accessible.}}

\begin{abstract}

InGaZnO (IGZO)-channel FeFETs have attracted notable interest thanks to recent advances in endurance, opening up their application space for read-dominated AI memory tiers. This work evaluates the viability of NOR-type IGZO FeFETs for 3D heterogeneous AI memories from a read-centric design-technology co-optimization (DTCO) perspective, spanning on-chip back-end-of-line (BEOL) RAMs and hybrid-bonded memory chiplets, and off-chip, monolithically integrated 3D FeNOR storage-class memories (SCMs). For on-chip BEOL RAMs and memory chiplets, we demonstrate the cross-node bitcell footprint scalability of IGZO FeFETs capable of delivering down to 10-\AA{} SRAM-equivalent bitcell area ($\sim$0.016 $\mu$m$^2$) with 7-nm ground rules while maintaining a sub-5 ns random access latency -- despite their writability challenges. We further identify the sensing margin penalty in NOR FeFET arrays arising from sneak current associated with the negative program-state $V_t$, which requires positive-$V_t$ engineering in order to eliminate the unwanted negative voltage read inhibition -- for example, by ferroelectric layer thinning. Last but not least, we elucidate the read margin implications on 3D FeNOR for SCMs, with the 3D stacking density limited by additional sneak current from neighbor channel shunting.

\end{abstract}

\begin{IEEEkeywords}
NOR FeFET, DTCO, BEOL RAM, Memory Chiplet, Storage Class Memory (SCM).
\end{IEEEkeywords}


\maketitle

\section{INTRODUCTION}
\label{sect1}

Oxide semiconductor channel (OSC) materials such as IGZO have been garnering renewed attention for FeFETs \cite{R1}\cite{R2}\cite{R3} thanks to their interfacial oxide-free properties that extend the cycling endurance to $>10^{12}$ \cite{R3}. Such breakthrough opens up the application space of FeFETs for read-intensive memories in large language model (LLM) inference, in an era where LLM growth is relentlessly driving up the demand for memory capacity and bandwidth \cite{R4}. In particular, the BEOL compatibility of OSC FeFETs along with their compact 1T bitcell \cite{R3} makes them a promising candidate for low-cost on-chip caching \cite{R5}, while recent reports on high-endurance monolithically integrated 3D OSC-FeFETs \cite{R6} additionally point to their potential as a low-voltage and high-endurance alternative to NAND flash-based low-latency SCMs \cite{R7}\cite{R8}.

\begin{figure}
  \centering
     \includegraphics[scale=0.57]{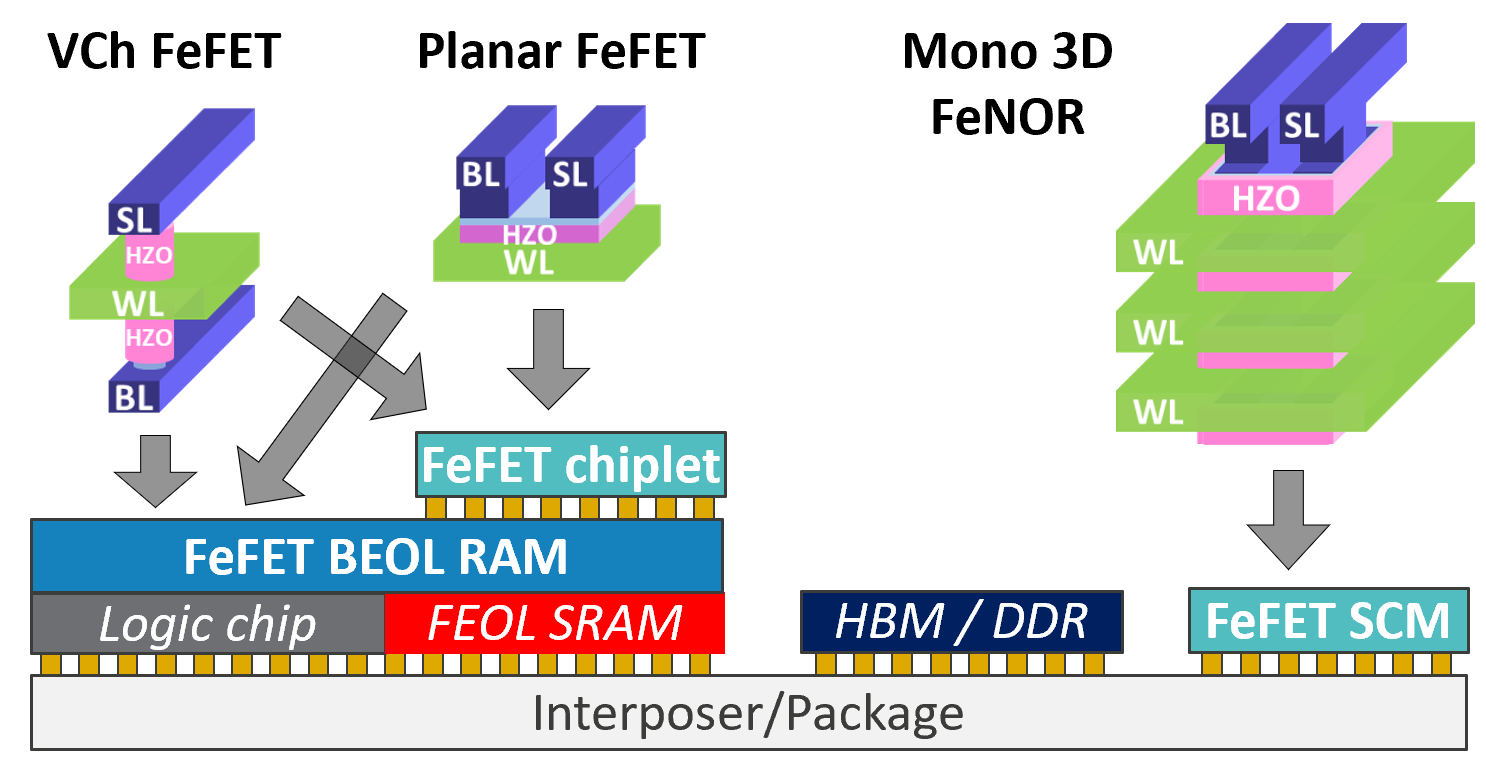}
\caption{OSC-FeFET application space in AI inference hardware: on-chip as BEOL RAM or hybrid bonded chiplet atop logic chip and FEOL SRAM, and off-chip as SCM connected to the logic chip via specialized protocols. Three types of FeFETs are considered: vertical-channel (VCh) FeFET (after \cite{R9}), planar FeFET (after \cite{R2}\cite{R3}) and monolithically integrated (“mono”) 3D FeNOR (after \cite{R6}).}
\label{F1}
\end{figure}

\begin{figure}
  \centering
     \includegraphics[scale=.94]{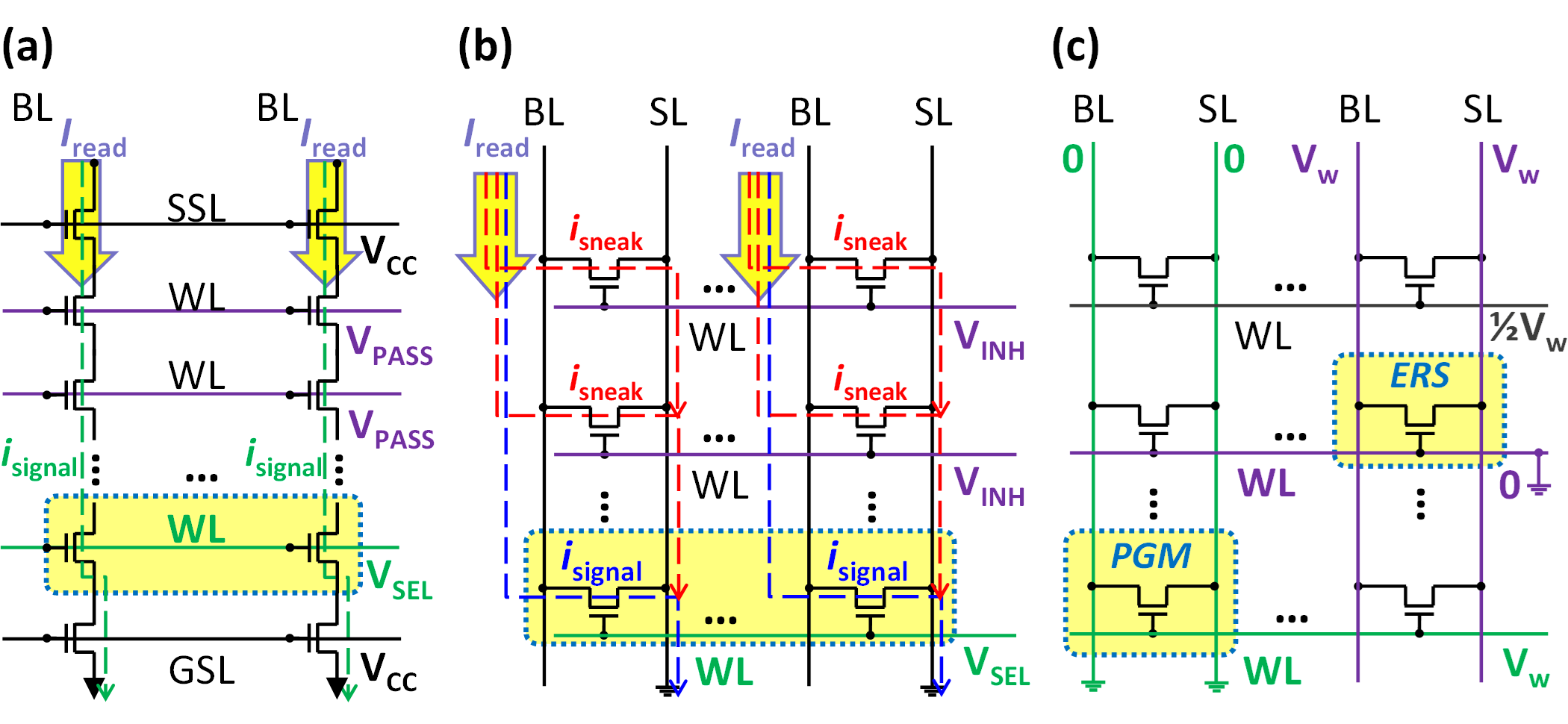}
\caption{Equivalent circuit schematic of (a) NAND FeFET array (read), (b)(c) NOR FeFET array in (b) read and (c) write (using the example of PGM). The read current in (a) NAND consists of the signal current ($i_{\mathrm{signal}}$) alone that flows serially through the string with all unselected rows biased at a pass voltage ($V_{\mathrm{PASS}}$), while in (b) NOR it parallelly sums up $i_{\mathrm{signal}}$ and all sneak current ($i_{\mathrm{sneak}}$) from unselected rows as a function of inhibition voltage ($V_{\mathrm{INH}}$). (c) NOR FeFET supports cross-point random write at $V_{w}$ (write) and $\frac{1}{2}V_{w}$ (inhibition), respectively.}
\label{F2}
\end{figure}

This work assesses the power-performance-area (PPA) scaling prospects of NOR-type IGZO FeFETs through comprehensive DTCO, targeting read-intensive AI memory use cases across compact on-chip BEOL RAMs and hybrid-bonded memory chiplets, as well as off-chip storage-class memories (Fig.~\ref{F1}). We specifically focus on AI inference workloads where the memory traffic in self-attention involves predominantly weights- and KV-cache reading at a read-to-write ratio up to $\sim 10^{3}$ \cite{R4}, which presumably eases FeFET write programming (PGM) / erasing (ERS) speed bottleneck ($> 100$ ns; \cite{R2}) compared to, e.g., SRAM-based cache ($< 10$ ns; \cite{R5}). For storage-class memories, we concentrate on the NOR configuration because it offers significantly lower read latency than NAND \cite{R10}, as its read current ($I_{\mathrm{read}}$) does not decrease with the BL string length, but instead becomes sensitive to accumulated sneak current ($i_{\mathrm{sneak}}$) from unselected rows (Fig.~\ref{F2}(a)(b)). Meanwhile, the random PGM/ERS capability of NOR-FeFETs (Fig.~\ref{F2}(c)) and speed advantages \cite{R11} compared with block-erase NAND FeFETs make them more suitable for AI inference scenarios with non-negligible update traffic, such as KV-cache generation during decoding \cite{R4}.

Overall, this paper aims to examine the viability of NOR IGZO FeFETs against the backdrop of heterogeneous AI memory tiers, with a particular emphasis on read-centric scaling constraints. The key question is not only to what extent compact FeFET bitcells improve memory bit density, but also whether sufficient read sensing margin and energy efficiency can be preserved as array size and 3D stacking grow. To this end, Section~\ref{sect2} discusses the bitcell scaling of on-chip FeFETs. Section~\ref{sect3} evaluates their read power-performance behavior, including the impact of NOR-array sneak current and programmed-state $V_t$ engineering. Section~\ref{sect4} extends the analysis to monolithically integrated (“mono”) 3D FeNOR storage-class memories, where channel shunting introduces an additional read-margin constraint. Section~\ref{sect5} summarizes the resulting device-, array-, and integration-level design implications.

\section{BITCELL SCALING OF BEOL AND CHIPLET FEFET MEMORIES}
\label{sect2}

\begin{figure}
  \centering
     \includegraphics[scale=.91]{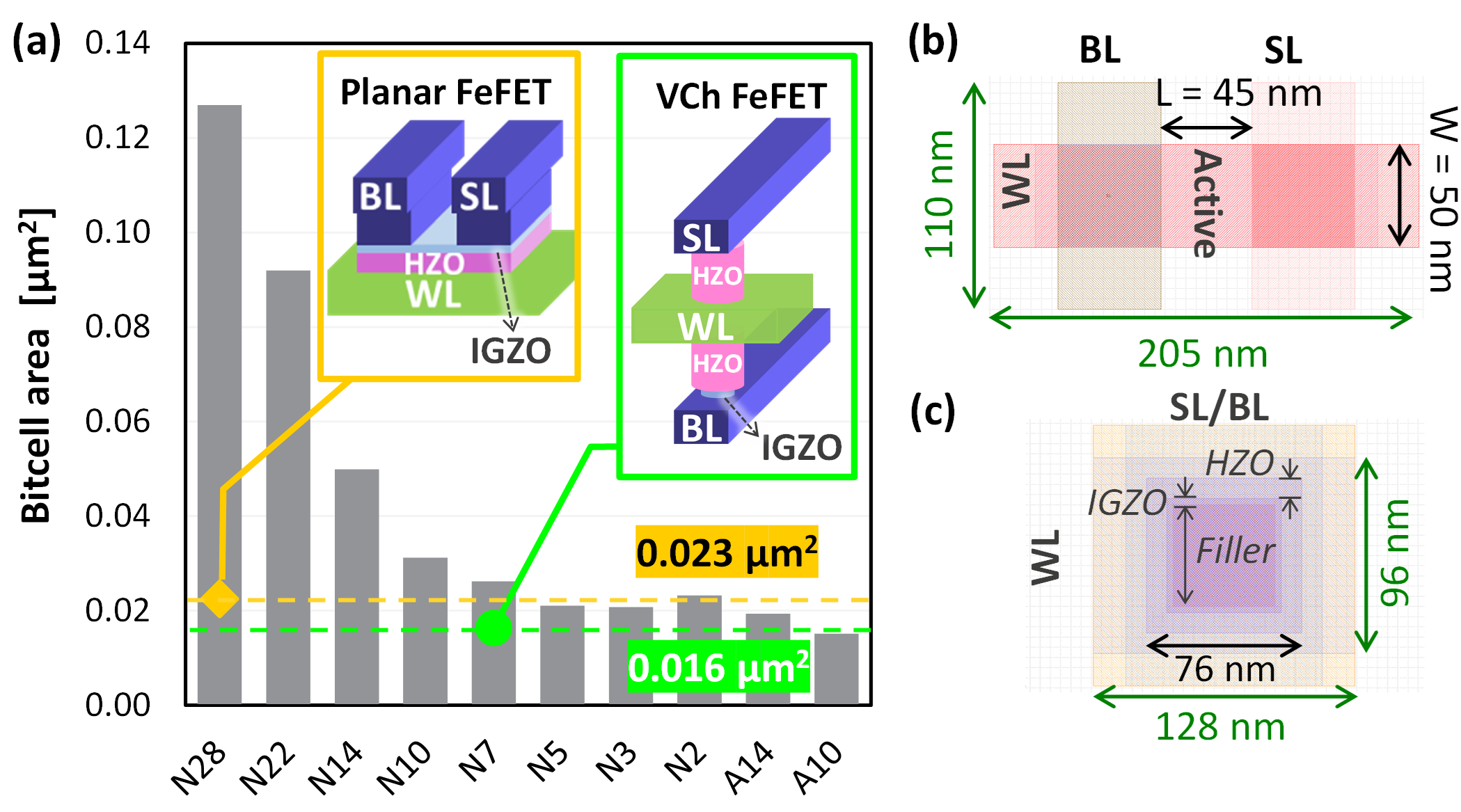}
\caption{(a) On-chip FeFET bitcell area versus SRAM scaling roadmap \cite{R12}, with planar and vertical-channel (VCh) FeFETs reaching sub-N2 and A10 SRAM footprint, respectively. (b)(c) Bitcell layouts for (b) planar and (c) VCh FeFETs, with the (b) planar FeFET based on dimensions in \cite{R3} (approximately 28 nm-node design rules) and (c) VCh FeFET whose WL/BL/SL are implemented in N7 BEOL M9.}
\label{F3}
\end{figure}

For on-chip applications we consider single-stack, planar \cite{R3} or vertical-channel (VCh) \cite{R9} IGZO-FeFETs in Fig.~\ref{F3}(a). We envision them to be either designed and routed on upper BEOL metal layers of advanced logic chips, as BEOL RAMs, or alternatively custom designed and hybrid bonded to logic as memory chiplets \cite{R12}. 
In particular, the VCh FeFET as a BEOL RAM can be implemented between three BEOL metal layers where the top/bottom metals serve as SL/BL, respectively while the middle layer defines the WL; the IGZO channel and (doped) Hf$_{0.5}$Zr$_{0.5}$O$_{2}$ (HZO) layers and the oxide filler are used to fill the “via” across the three metals, making it a $4F^{2}$ bitcell (Fig.~\ref{F3}(a)(c)).
In both cases, the 1T compact-bitcell design accords on-chip FeFETs significant “cross-node” bitcell footprint scaling versus SRAM-based alternatives (Fig.~\ref{F3}(a)) that have turned increasingly challenging/costly to scale in beyond-FinFET era \cite{R13}. Indeed, the planar FeFET bitcell based on latest hardware dimensions (\cite{R3}; $\sim 28$ nm design rules) easily outscales N2 SRAM (0.023 $\mu$m$^{2}$), whereas the VCh FeFET using N7 M9 BEOL layer further pushes the bitcell area towards 10 Å SRAM (0.016 $\mu$m$^{2}$) \cite{R13}. Additional tightening of planar and/or VCh FeFET design rules is expected to deliver even more scalable and cost-effective on-chip memory functionality to read-dominated workloads.


\section{READ POWER-PERFORMANCE OF VERTICALLY INTEGRATED ON-CHIP FEFET MEMORIES}
\label{sect3}

\begin{figure}
  \centering
     \includegraphics[scale=.94]{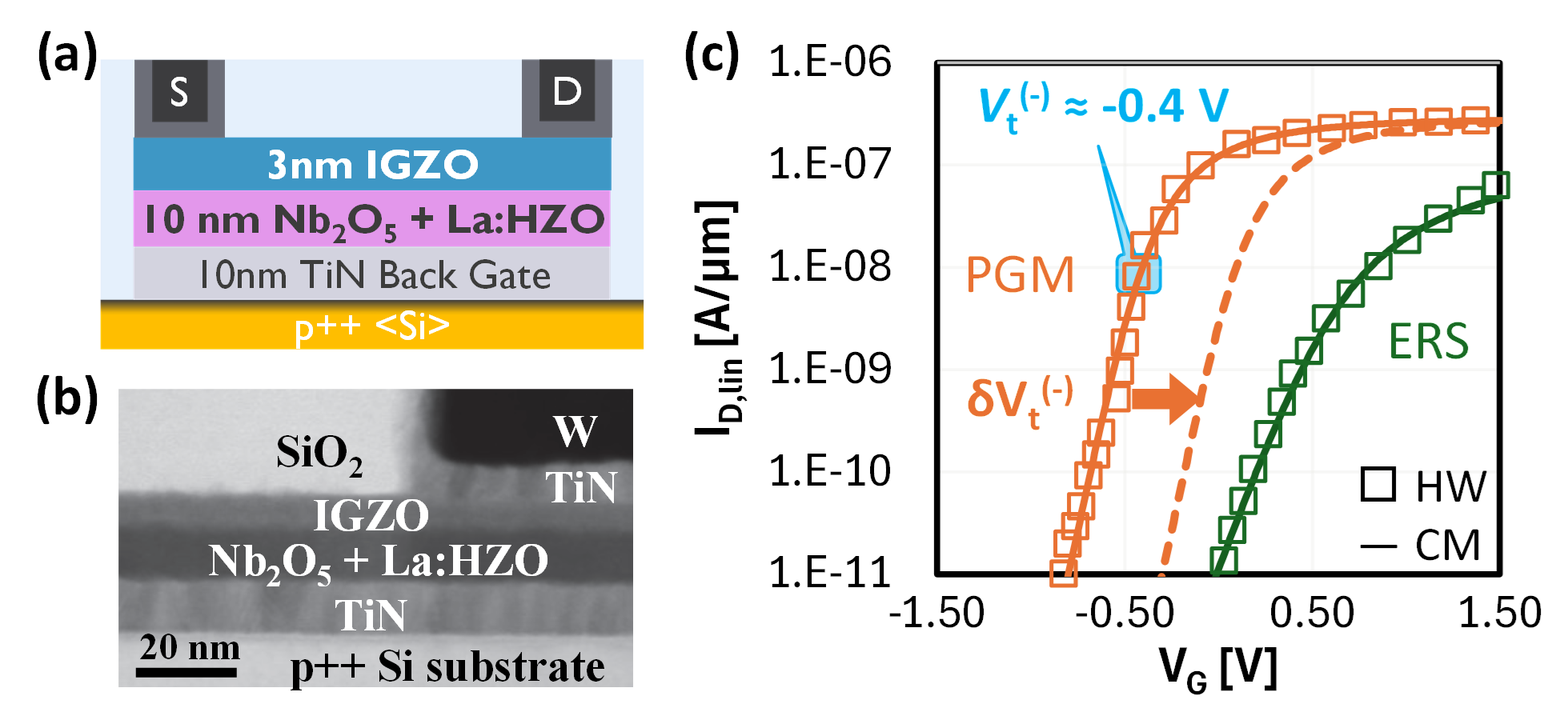}
\caption{(a) Schematic and (b) TEM cross section of planar FeFET device in \cite{R2}. (c) BSIM-IMG compact model \cite{R17} fitted to $I$-$V$ measurements on hardware in \cite{R2} (25 $^\circ$C; $L_{g} \approx 130$ nm). The $V_{t}^{(-)}$ sits at around $-0.4$ V. Simulations are performed based on or extrapolated from calibrated CM (e.g., with artificially shifted $V_{t}^{(-)}$).}
\label{F4}
\end{figure}

Despite the perceived read speed advantage of NOR-FeFETs, a commonly encountered read challenge is the sneak current ($i_{\mathrm{sneak}}$) from the same column that is indiscriminately captured in $I_{\mathrm{read}}$ (Fig.~\ref{F2}(b)), which could hinder the readability of selected words \cite{R14}. To assess the viability of proposed on-chip FeFETs in Fig.~\ref{F3}, we perform SPICE \cite{R15} array simulations with layout-related parasitics extracted using TCAD \cite{R16}. The $I$-$V$ characteristics are captured by the BSIM-IMG \cite{R17} compact model (CM) calibrated to hardware (HW) measurements in \cite{R2} (Fig.~\ref{F4}) and extrapolated per dimensional scaling of BSIM-IMG. In addition, we allow artificial shifting of the PGM-state $V_{t}$ (referred to as $V_{t}^{(-)}$) to probe its sensitivity in reading. The following analysis therefore treats read margin as the central constraint linking device design to memory-tier scalability. In a NOR array, the selected-cell signal is sensed together with current contributions from unselected rows (see Fig.~\ref{F2}(b)), so the programmed-state $V_t$, inhibition bias, and array size directly determine both read robustness and energy.

\subsection{\textsc{FeFET array read sensitivity to PGM state $V_t$}}
\label{sect3A}

\begin{figure}
  \centering
     \includegraphics[scale=.94]{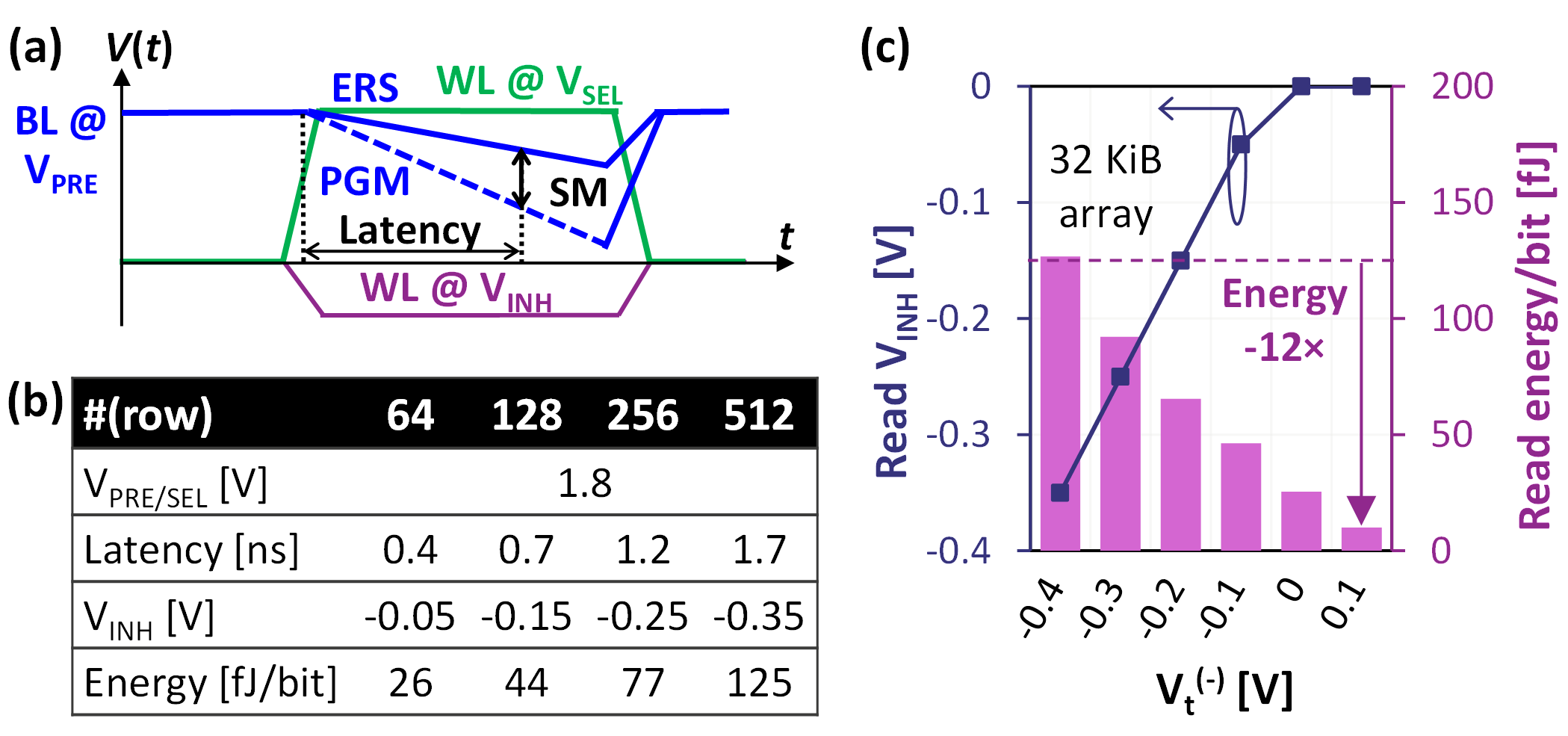}
\caption{(a) Read timing diagram for FeFET NOR array, where the BL is first precharged to $V_{\mathrm{PRE}}$ and then discharges based on FeFET read current in PGM or ERS states. WL $V_{\mathrm{SEL}}$ is set equal to $V_{\mathrm{PRE}}$. A negative $V_{\mathrm{INH}}$ is applied as appropriate. For simplicity, the read latency and energy do not include comparator latching. (b) Array read metrics for planar FeFET in Fig.~\ref{F3}(b) as a function of number of rows; successful read is recorded only when a sensing margin (SM) of more than 100 mV is reached. (c) Planar FeFET sensitivity of required $V_{\mathrm{INH}}$ and read energy in a 32 KiB array vs shifted $V_{t}^{(-)}$ in Fig.~\ref{F4}(c).}
\label{F5}
\end{figure}

As evidenced in simulations (Fig.~\ref{F5}(b)), the NOR-FeFET array (here using planar without loss of generality) indeed allows for ultra-low read delay (sub-ns) even by using the simplest BL discharging read mechanism (Fig.~\ref{F5}(a)), thanks to the parallel connection of all bitcells on the same BL (Fig.~\ref{F2}(b)). However, the accumulation of parallel $i_{\mathrm{sneak}}$ in the array with increasing number of rows – especially in the extreme case when all $i_{\mathrm{sneak}}$ paths are produced by the negative $V_{t}^{(-)}$ (Fig.~\ref{F4}) – would swamp the actual $i_{\mathrm{signal}}$ in selection (Fig.~\ref{F2}(b)), such that a negative inhibition voltage ($V_{\mathrm{INH}}$) needs to be imposed on unselected WLs. Consequently, as the array size grows, so does the magnitude of $V_{\mathrm{INH}}$ for safeguarding a 100-mV BL read sensing margin (SM; Fig.~\ref{F5}(a)(b)). This not only introduces extra read energy consumption but also adds to circuital complexity in on-chip negative voltage generation \cite{R18}.

A countermeasure to the unwanted negative $V_{\mathrm{INH}}$, especially in large NOR arrays, would be to shift the FeFET $V_{t}^{(-)}$ to positive (Fig.~\ref{F4}; \cite{R14}), thus eliminating the circuit and energy overhead needed to curb $i_{\mathrm{sneak}}$; this of course must not compromise the robustness of memory window (MW), namely the difference between ERS- and PGM-state $V_{t}$’s ($V_{t}^{(\pm)}$) under variability \cite{R19}. The effect of such “virtual $V_{t}$ engineering” is shown in Fig.~\ref{F5}(c) for a 32 KiB array, where a positive $V_{t}^{(-)}$ would indeed remove the negative $V_{\mathrm{INH}}$ requirement, with the read energy reduced by 12$\times$ at 0.1 V $V_{t}^{(-)}$. A positive $V_{t}^{(-)}$ is therefore highly desirable. This result establishes programmed-state $V_t$ as a device-level knob with direct array-level consequences for read biasing complexity and energy consumption in NOR-type IGZO FeFETs.

\subsection{\textsc{FeFET $V_t$ engineering by FE layer thinning}}
\label{sect3B}

In this subsection, we discuss the possibility of FE layer thickness ($t_{\mathrm{FE}}$) scaling for implementing the positive-$V_{t}$ engineering proposed in Section~\ref{sect3}-\ref{sect3A}, using known FeFET $V_{t}$ relations established in \cite{R20} while making necessary adaptations to the metal-FE-IGZO (MFS) structure of IGZO-FETs (Fig.~\ref{F4}(a)). For simplicity, we ignore any oxide bulk or interface defects.

\begin{figure}
  \centering
     \includegraphics[scale=.40]{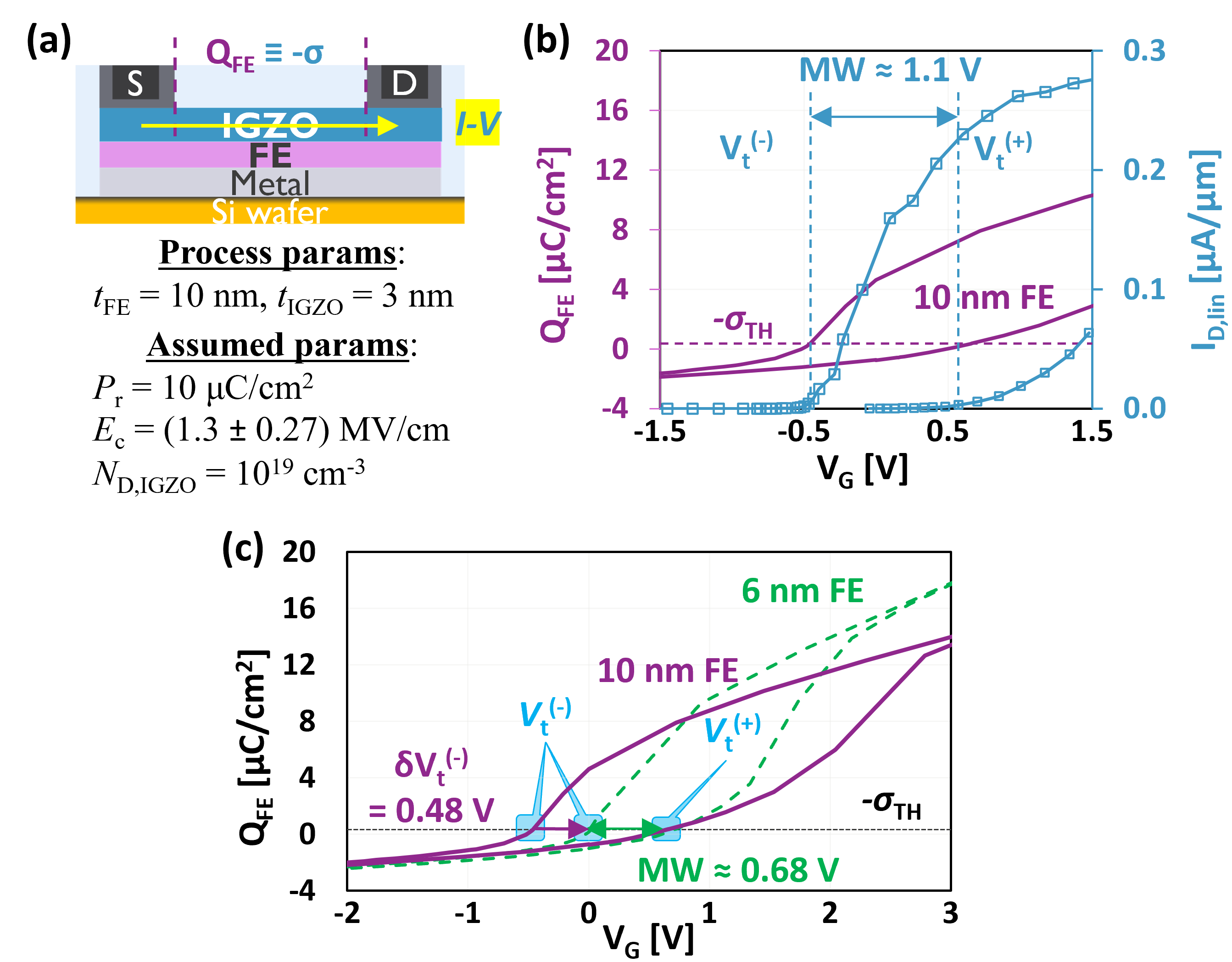}
\caption{(a) Metal-FE-IGZO (MFS) stack in IGZO-FeFET where $Q_{\mathrm{FE}}$ in FE and $-\sigma$ in IGZO continue across their interface. (b) Simulated $Q_{\mathrm{FE}}$-$V_{G}$ characteristics using parameters in (a) that match the measured $V_{t}^{(\pm)}$ (from $I$-$V$) at the loadline intersections $Q_{\mathrm{FE}} = -\sigma_{\mathrm{TH}}$ (here approx. $0.35~\mu$C/cm$^{2}$ or $2.2\times10^{12}$ cm$^{-2}$). (c) Extrapolation of $Q_{\mathrm{FE}}$-$V_{G}$ characteristic to 6 nm thick FE layer, with the extrapolated $V_{t}^{(\pm)}$ corresponding to the intercepts of $Q_{\mathrm{FE}}$-$V_{G}$ with the loadline $Q_{\mathrm{FE}} = -\sigma_{\mathrm{TH}}$. The $V_{t}^{(-)}$ is shifted by 0.48 V when $t_{\mathrm{FE}}$ scales from 10 nm to 6 nm.}
\label{F6}
\end{figure}

First, we rewrite the FeFET $V_{t}$ equation \cite{R20} as
\begin{equation}
V_{t}^{(\pm)} = V_{\mathrm{FB}} + \psi_{s,\mathrm{TH}} + t_{\mathrm{FE}} E_{\mathrm{FE,TH}}^{(\pm)}\,,
\label{E1}
\end{equation}
where $\psi_{s,\mathrm{TH}}$ and $E_{\mathrm{FE,TH}}$ refer to the channel surface potential and the electric field across FE layer at threshold; the $\pm$ sign stands for ERS/PGM states, respectively. The $E_{\mathrm{FE,TH}}$ is in turn given by Eq. (\ref{E2}) per\cite{R20}:
\begin{equation}
-\sigma_{\mathrm{TH}} = \epsilon_{\mathrm{FE}} E_{\mathrm{FE,TH}}^{(\pm)} + P_{\mathrm{FE}}^{(\pm)}\big|_{\sigma = \sigma_{\mathrm{TH}}} \equiv {Q_{FE}}\big|_{\sigma = \sigma_{\mathrm{TH}}}\,,
\label{E2}
\end{equation}
with $\sigma_{\mathrm{TH}}$ being the IGZO channel charge density at threshold. Noting $Q_{\mathrm{FE}} \equiv -\sigma$ across IGZO-FE interface (Fig.~\ref{F6}(a)), $E_{\mathrm{FE,TH}}$ is evidently the cross point(s) between $Q_{\mathrm{FE}}$-$E_{\mathrm{FE}}$ hysteresis and $\sigma \equiv -Q_{\mathrm{FE}} = \sigma_{\mathrm{TH}}$. Combining $V_{\mathrm{FB}}$ and $\psi_{s}$ as one IGZO voltage ($V_{\mathrm{IGZO}}$):
\begin{equation}
V_{t}^{(\pm)} = V_{\mathrm{IGZO}} \big|_{Q_{\mathrm{FE}} = -\sigma_{\mathrm{TH}}} + t_{\mathrm{FE}} \times E_{\mathrm{FE}}^{(\pm)} \big|_{Q_{\mathrm{FE}} = -\sigma_{\mathrm{TH}}}\,.
\label{E3}
\end{equation}
The right-hand side of Eq. (\ref{E3}) essentially assigns FeFET $V_{t}^{(\pm)}$ to the cross points between the $Q_{\mathrm{FE}}$-$V_{G}$ hysteresis in the MFS structure in Fig.~\ref{F6}(a) and the “loadline” $Q_{\mathrm{FE}} = -\sigma_{\mathrm{TH}}$. Indeed, as evidenced in Fig.~\ref{F6}(b), such relationship is reasonably matched by using known MFS stack parameters and common assumptions for IGZO-FeFETs \cite{R21}.

\begin{figure}
  \centering
     \includegraphics[scale=.48]{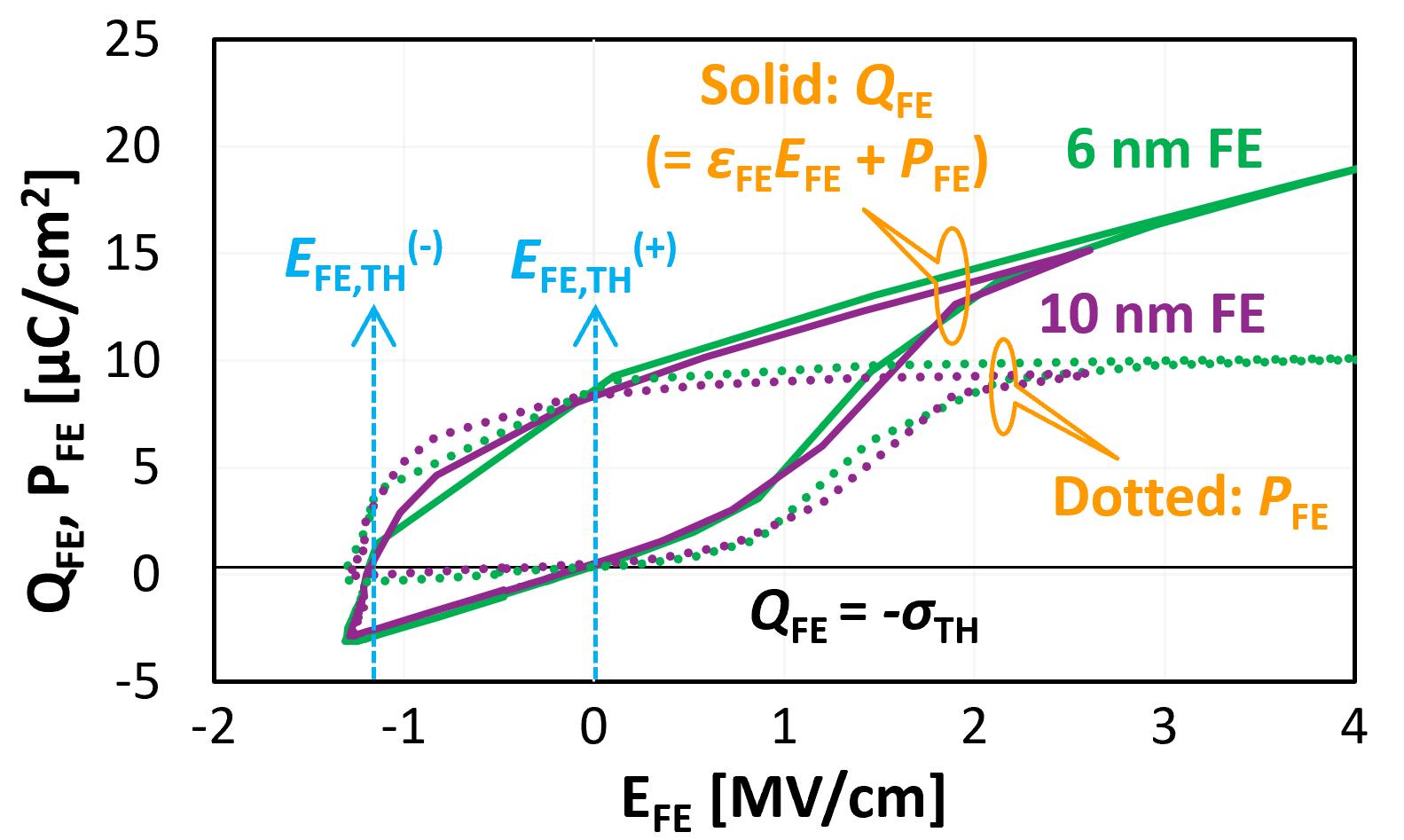}
\caption{Simulated $Q_{\mathrm{FE}}$-$E_{\mathrm{FE}}$ (solid) and $P_{\mathrm{FE}}$-$E_{\mathrm{FE}}$ (dotted) hystereses for the 10 nm and 6 nm FE-layer-based MFS stacks in Fig.~\ref{F6}(c).}
\label{F7}
\end{figure}

Further explorations (Fig.~\ref{F6}(c)) based on Eq. (\ref{E3}) show that by shrinking $t_{\mathrm{FE}}$ to 6 nm \cite{R22}, one may indeed achieve the anticipated positive $\delta V_{t}^{(-)}$ as in Fig.~\ref{F4}(c). The “single-sided” $V_{t}$ shift can be explained by the fact that the $Q_{\mathrm{FE}}$-$V_{G}$ hystereses in Fig.~\ref{F6}(c) are in fact subloops where full erase (i.e., $P_{\mathrm{FE}} < 0$) cannot be effectively achieved due to the lack of compensating positive hole carriers in IGZO \cite{R2}, for which $Q_{\mathrm{FE}}$-$V_{G}$ appears “pinched off” below $Q_{\mathrm{FE}} = 0$. Consequently, the $E_{\mathrm{FE,TH}}^{(+)}$ in Eq. (\ref{E3}) is almost zero as $P_{\mathrm{FE}}$ is hardly ever negative, whereas $E_{\mathrm{FE,TH}}^{(-)}$ that corresponds to the positive half of $Q_{\mathrm{FE}}$-$V_{G}$ has traversed a significant part of the positive $P_{\mathrm{FE}}$-$V_{G}$ hysteresis and hence deviates notably from 0 (Fig.~\ref{F7}). In sum, we confirm the theoretical feasibility of achieving positive $V_{t}^{(-)}$ by $t_{\mathrm{FE}}$ scaling.

\subsection{\textsc{Benchmarking of on-chip IGZO FeFETs vs. SRAM}}
\label{sect3C}

Considering the planar and VCh FeFET bitcell layout options in Section~\ref{sect2} and the readability analyses in Section~\ref{sect3}-\ref{sect3A}, we here compare the full PPA of on-chip FeFETs versus advanced SRAM \cite{R13} for read-centric workloads (Table~\ref{Tab1}). For completeness, preliminary device write specifications are also included based on \cite{R2}. Clearly, the projected negative inhibition-lifting positive-$V_{t}^{(-)}$ device knob would be instrumental in reducing the read energy consumption of both planar and VCh FeFETs, bringing it down to a similar level of A10 SRAM (max. +35 \%) while sustaining sub-5 ns read latency and virtually no leakage. This, plus the cross-node bitcell area scaling (cf. Fig.~\ref{F3}(a)) makes on-chip FeFETs, especially BEOL VCh FeFETs, particularly appealing for enhancing on-chip memory capacity and locality in read-dominated workloads. Meanwhile, further device tuning is expected to reduce the write voltage and time in FeFETs (e.g., $< 30$ ns at $\pm 2.4$ V \cite{R3}) to allow for more efficient writing, e.g., KV cache appending during inference decode \cite{R4}.

\begin{table}[t]
\caption{On-Chip FeFET PPA versus A10 SRAM (32 KiB; 25 $^\circ$C)}
\label{Tab1}

\setlength{\tabcolsep}{1.5pt}

\resizebox{\columnwidth}{!}{%
\begin{tabular}{p{78pt}p{42pt}p{42pt}p{42pt}p{42pt}p{42pt}}

\hline

&
\multicolumn{2}{c}{Planar FeFET} &
\multicolumn{2}{c}{VCh FeFET} &
A10 SRAM \cite{R13}
\\

\cline{2-6}

&
$V_{t}^{(-)}$

@$-0.4$ V &
$V_{t}^{(-)}$

@$+0.1$ V &
$V_{t}^{(-)}$

@$-0.4$ V &
$V_{t}^{(-)}$

@$+0.1$ V &
\\

\hline

Bitcell area [$\mu$m$^{2}$]
&
0.023
&
\multicolumn{3}{c}{0.016}
&
0.015
\\

Read voltage [V]
&
1.8
&
1.4
&
1.8
&
1.4
&
0.7
\\

Inhibition voltage [V]
&
-0.35
&
0
&
-0.35
&
0
&
N/A
\\

Read delay [ns]
&
1.7
&
1.8
&
2.9$^{b}$
&
4.4$^{c}$
&
$< 0.3$
\\

Read energy [fJ/bit]
&
125
&
10.1
&
126
&
7.5
&
8.4
\\

Leakage power [$\mu$W]
&
\multicolumn{4}{c}{0}
&
158
\\

Write voltage [V]
&
\multicolumn{4}{c}{$\pm 3.5$ \cite{R2}}
&
0.7
\\

Write delay [ns]
&
\multicolumn{4}{c}{$> 100$ \cite{R2}}
&
$< 0.3$
\\

\hline

\multicolumn{6}{p{288pt}}{
$^{a}$ Electrical characteristics of which are scaled from those of planar FeFET in Fig.~\ref{F4}(c).
}
\\

\multicolumn{6}{p{288pt}}{
$^{b,c}$ VCh FeFET is slower due to a longer gate defined by WL thickness, i.e., M9 thickness (160 nm).
}
\\

\end{tabular}
}
\end{table}

\section{SCALABILITY OF MONOLITHIC 3D NOR FEFET-BASED STORAGE CLASS MEMORIES}
\label{sect4}

Another read-dominated AI memory application space we envision for FeFETs is the high-capacity, high-throughput storage-class memories (Fig.~\ref{F1}) that are often either 2.5D co-packaged \cite{R7} or standalone \cite{R8}. Process integration-wise this has been demonstrated with a monolithic 3D stacking process flow for 3D FeNORs similar to that for 3D NAND flash, albeit aimed at in-memory computing \cite{R6}\cite{R11}. The PPA scalability of SCM-centric 3D FeNOR, in the meantime, has yet to be verified. This section therefore extends the same read-centric DTCO perspective from planar array scaling to stack-level scalability.

\begin{figure}
  \centering
     \includegraphics[scale=.94]{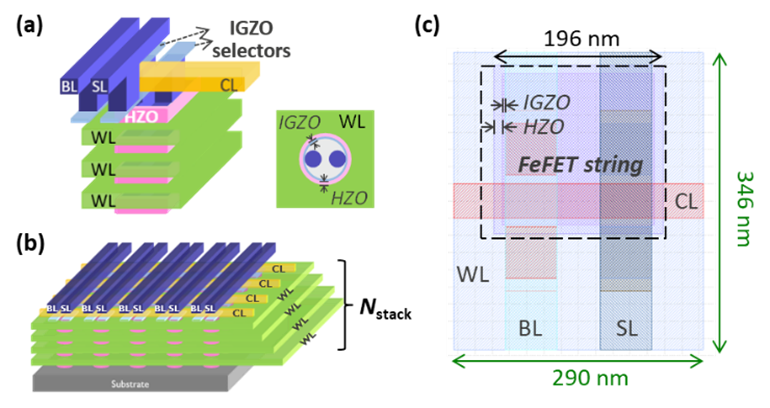}
\caption{(a) Mono 3D FeNOR string and (b) array structure, with a cross section of the string (at WL plane) given in the inset of (a). IGZO-FET \cite{R23}-based BL and SL selectors are included per string, sharing the gate control line (CL). (c) Bitcell layout for one string including BL and SL IGZO-FET selectors.}
\label{F8}
\end{figure}

Building on our DTCO methodology in Sections~\ref{sect2} and~\ref{sect3}, we show in Fig.~\ref{F8} a layout option of 3D FeNOR SCM assuming similar monolithic integration processes in \cite{R6}\cite{R11}. Notably, for SCM that requires word selection, separate string selectors are a prerequisite for discriminating different strings on the same BL/SL. This is implemented using a pair of planar IGZO FETs \cite{R23} with shared gate control line (CL) per string, for BL and SL, respectively. The extra footprint of the selectors is to be compensated by maximizing vertical stacking ($N_{\mathrm{stack}}$).

\begin{figure}
  \centering
     \includegraphics[scale=.61]{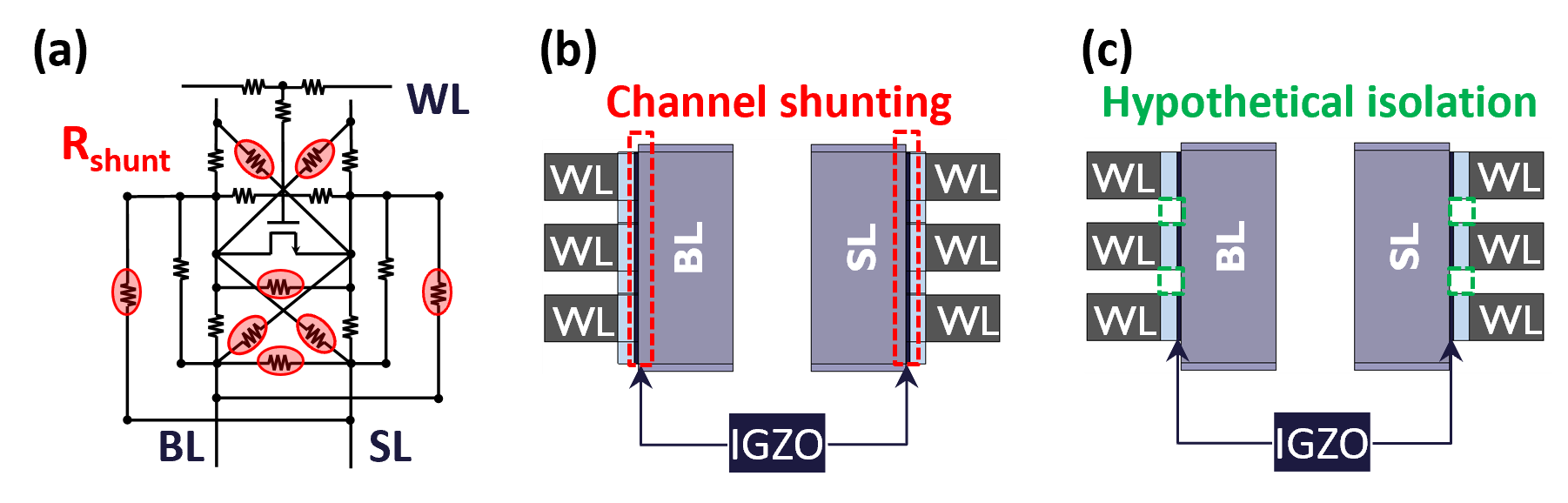}
\caption{(a) Equivalent resistance circuit extracted by TCAD \cite{R16} indicating shunting resistance ($R_{\mathrm{shunt}}$) in ungated regions of 3D FeNOR between neighboring FeFETs on the same vertical string due to (b) contiguous IGZO layer across different WL planes formed in monolithic integration. (c) A hypothetical version of 3D FeNOR where isolation is placed in ungated regions.}
\label{F9}
\end{figure}

In maximizing $N_{\mathrm{stack}}$ in monolithic vertical stacking, it is established through TCAD parasitic extraction \cite{R16} that there is extra contribution to $i_{\mathrm{sneak}}$ (Fig.~\ref{F2}(b)) in 3D FeNOR structure due to the contiguous IGZO channel conformally deposited in string formation \cite{R6}\cite{R11}. It creates a shunted path across different WL planes as finite resistance ($R_{\mathrm{shunt}}$) in ungated regions between WL planes on the same string (Fig.~\ref{F9}(a)(b)), which imposes, on top of the negative $V_{t}^{(-)}$ (Section~\ref{sect3}-\ref{sect3A}), additional penalty on the readability in highly stacked strings.

The effect of channel shunting is confirmed by SPICE simulations in a 2 KiB array (Table~\ref{Tab2}): we observe that with channel shunting, the maximum $N_{\mathrm{stack}}$ can hardly exceed four for a minimum 100-mV SM – even with the proposed positive $V_{t}^{(-)}$ implemented (Sections~\ref{sect3}-\ref{sect3A} and \ref{sect3}-\ref{sect3B}) – which limits the bit density of 3D FeNOR to only $\sim 28$ Mb/mm$^{2}$ that is in SRAM range \cite{R24}. In contrast, in a 3D FeNOR structure that allows isolation in ungated regions (Fig.~\ref{F9}(c)), the 100 mV-SM allows $N_{\mathrm{stack}}$ to go up to 64 (under the +0.1 V $V_{t}^{(-)}$ assumption in Section~\ref{sect3}-\ref{sect3A}) and accordingly the bit density up to 0.45 Gb/mm$^{2}$ that exceeds advanced 1$\beta$ DRAM \cite{R25}. It is therefore essential that proper channel isolation techniques be developed in order to ensure the electrical functionality of 3D FeNOR-based SCMs.

\begin{table}[t]
\caption{Simulated Read PPA of 3D FeNOR (2 KiB array; 25 $^\circ$C)$^{a}$}
\label{Tab2}

\setlength{\tabcolsep}{1.5pt}

\resizebox{\columnwidth}{!}{%
\begin{tabular}{p{95pt}p{42pt}p{42pt}p{42pt}p{42pt}}

\hline

&
\multicolumn{2}{c}{Finite $R_{\mathrm{shunt}}$}
&
\multicolumn{2}{c}{Hypothetical $R_{\mathrm{shunt}} = \infty$}
\\

\cline{2-5}

&
$V_{t}^{(-)}$

@ $-0.4$ V
&
$V_{t}^{(-)}$

@$+0.1$ V
&
$V_{t}^{(-)}$

@
$-0.4$ V
&
$V_{t}^{(-)}$

@
$+0.1$ V
\\

\hline

Bitcell area [$\mu$m$^{2}$]
&
\multicolumn{4}{c}{0.10}
\\

Vertical WL pitch [nm]$^{b}$
&
\multicolumn{4}{c}{45}
\\

$N_{\mathrm{stack}}$ under SM $> 100$ mV
&
\multicolumn{2}{c}{4}
&
16
&
64
\\

Bit density [Gb/mm$^{2}$]$^{c}$
&
\multicolumn{2}{c}{0.028}
&
0.11
&
0.45
\\

Read voltage [V]
&
1.8
&
1.7
&
1.8
&
1.2
\\

Inhibition voltage [V]
&
-0.5
&
0
&
-0.35
&
0
\\

Read delay [ns]
&
0.7
&
0.8
&
0.4
&
1.0
\\

Read energy [pJ/bit]
&
1.48
&
1.30
&
1.44
&
0.62
\\

\hline

\multicolumn{5}{p{263pt}}{
$^{a}$ Electrical characteristics of which are scaled from those of planar FeFET in Fig.~\ref{F4}(c).
}
\\

\multicolumn{5}{p{263pt}}{
$^{b}$ Including 30 nm WL thickness and 15 nm spacer.
}
\\

\multicolumn{5}{p{263pt}}{
$^{c}$ Assume $\sim 70$ \% area efficiency \cite{R14}.
}
\\

\end{tabular}
}
\end{table}

\section{CONCLUSION}
\label{sect5}

This paper presented a read-centric DTCO study of NOR-type IGZO FeFETs for 3D heterogeneous AI memories. The central finding is that their scalability is governed not only by bitcell footprint, but also by the preservation of read sensing margin in NOR arrays and for 3D FeNOR, additionally in vertically stacked configurations. For on-chip memories, we highlighted the cross-node bitcell footprint scalability of both planar and vertical channel FeFETs as BEOL RAMs and/or memory chiplets, that reach N2 to A10 SRAM bitcell area while maintaining relaxed, low-cost technology ground rules (N28 to N7); in the meantime they were electrically projected to be capable of achieving sub-5 ns random access with comparable energy consumption to advanced SRAMs in reading. For storage-class memories, we demonstrated sub-ns read latency in 3D FeNOR but also identified the neighbor shunting in monolithically deposited IGZO channel as a major constraint on read sensing margin when pursuing maximum 3D stacking, which needs to be overcome in order to unlock the bit density scaling potential of 3D FeNOR to 1$\beta$-DRAM level (0.45 Gb/mm$^{2}$). Overall, the results identify positive programmed-state $V_t$ engineering and channel isolation as the two key technology knobs for extending NOR IGZO FeFETs across heterogeneous AI memory tiers, while further write-voltage and write-time scaling remains important for workloads with non-negligible update traffic.

\section*{REFERENCES}

\def\refname{\vadjust{\vspace*{-1em}}} 


\begin{thebibliography}{00}

\bibitem{R1}
F. Mo \textit{et al.}, ``Experimental Demonstration of Ferroelectric HfO$_2$ FET with Ultrathin-body IGZO for High-Density and Low-Power Memory Application,'' \textit{2019 Symposium on VLSI Technology}, Kyoto, Japan, 2019, pp. T42--T43, doi: 10.23919/VLSIT.2019.8776553.

\bibitem{R2}
Z. Chen \textit{et al.}, ``Novel Design Strategy for High-Endurance ($>10^{10}$) and Fast-Erase Oxide-Semiconductor Channel FeFET,'' \textit{2024 IEEE International Electron Devices Meeting (IEDM)}, San Francisco, CA, USA, 2024, pp. 1--4, doi: 10.1109/IEDM50854.2024.10873449.

\bibitem{R3}
C.-C. Lu \textit{et al.}, ``Demonstration of Ferroelectric FET Memory with Oxide Semiconductor Channel to Achieve Smallest Cell Area 0.009 $\mu$m$^{2}$ and High Endurance for Non-Volatile High-Bandwidth Memory Applications,'' \textit{2024 IEEE International Electron Devices Meeting (IEDM)}, San Francisco, CA, USA, 2024, pp. 1--4, doi: 10.1109/IEDM50854.2024.10873402.

\bibitem{R4}
S. Legtchenko \textit{et al.}, ``Managed-Retention Memory: A New Class of Memory for the AI Era,'' arXiv:2501.09605v1 [cs.AR], 2025, doi: 10.48550/arXiv.2501.09605.

\bibitem{R5}
A. Lu, J. Lee, T. H. Kim \textit{et al.}, ``High-speed emerging memories for AI hardware accelerators,'' \textit{Nat. Rev. Electr. Eng.}, vol. 1, no. 1, pp. 24--34, 2024, doi: 10.1038/s44287-023-00002-9.

\bibitem{R6}
Y. Zhou \textit{et al.}, ``3D NOR-Type FeFETs with Record Endurance of $10^{11}$, Fast Erase of 50 ns, and Immediate Read-After-Write for In-Memory Learning,'' \textit{2025 Symposium on VLSI Technology and Circuits}, Kyoto, Japan, 2025, pp. 1--3, doi: 10.23919/VLSITechnologyandCir65189.2025.11074820.

\bibitem{R7}
Sandisk, ``Sandisk Unveils The Future of Memory Architecture for AI -- Introducing: High Bandwidth Flash,'' \textit{Tech Brief HBF Fact Sheet}, July 2025. [Online]. Available: \url{https://documents.sandisk.com/content/dam/asset-library/en_us/assets/public/sandisk/collateral/company/Sandisk-HBF-Fact-Sheet.pdf}

\bibitem{R8}
Kioxia, ``Opening New Doors in the Big Data Era: Low-Latency Flash as a Catalyst for Innovation,'' \textit{XL-FLASH Infographic}, 2025. [Online]. Available: \url{https://europe.kioxia.com/content/dam/kioxia/shared/business/memory/xlflash/asset/KIOXIA_XL-FLASH_Infographic.pdf}

\bibitem{R9}
J. Duan \textit{et al.}, ``A Full Spectrum of 3D Ferroelectric Memory Architectures Shape by Polarization Sensing,'' arXiv:2504.09713v1 [cs.ET], 2025, doi: 10.48550/arXiv.2504.09713.

\bibitem{R10}
H.-T. Lue \textit{et al.}, ``3D AND: A 3D Stackable Flash Memory Architecture to Realize High-Density and Fast-Read 3D NOR Flash and Storage-Class Memory,'' \textit{2020 IEEE International Electron Devices Meeting (IEDM)}, San Francisco, CA, USA, 2020, pp. 6.4.1--6.4.4, doi: 10.1109/IEDM13553.2020.9372101.

\bibitem{R11}
Y. Feng \textit{et al.}, ``First Demonstration of BEOL-Compatible 3D Vertical FeNOR,'' \textit{2024 IEEE Symposium on VLSI Technology and Circuits}, Honolulu, HI, USA, 2024, pp. 1--2, doi: 10.1109/VLSITechnologyandCir46783.2024.10631352.

\bibitem{R12}
A. Sharma \textit{et al.}, ``IGZO Based eDRAM: Bitcell and Array Optimization Enabling Denser Last Level Caches,'' \textit{2025 IEEE European Solid-State Electronics Research Conference (ESSERC)}, Munich, Germany, 2025, pp. 45--48, doi: 10.1109/ESSERC66193.2025.11214068.

\bibitem{R13}
D. Abdi \textit{et al.}, ``SRAM Scaling Opportunities Below 0.01 $\mu$m$^{2}$ Using Double-Row CFET Architecture with Wordline-Folded Bitcell Design for Performance Optimization,'' \textit{2025 Symposium on VLSI Technology and Circuits}, Kyoto, Japan, 2025, pp. 1--3, doi: 10.23919/VLSITechnologyandCir65189.2025.11075033.

\bibitem{R14}
T.-E. Lee \textit{et al.}, ``High-Endurance MoS$_2$ FeFET with Operating Voltage Less Than 1 V for eNVM in Scaled CMOS Technologies,'' \textit{2023 International Electron Devices Meeting (IEDM)}, San Francisco, CA, USA, 2023, pp. 1--4, doi: 10.1109/IEDM45741.2023.10413873.

\bibitem{R15}
\textit{Spectre Circuit Simulator}, Cadence, San Jose, CA, Mar. 2025.

\bibitem{R16}
\textit{Raphael FX User Guide}, Synopsys, Mountain View, CA, Dec. 2022.

\bibitem{R17}
G. Pahwa \textit{et al.}, \textit{BSIM-IMG 102.9.6 Independent Multi-Gate MOSFET Compact Model -- Technical Manual}, 2022. [Online]. Available: \url{https://www.bsim.berkeley.edu/models/bsimimg/}.

\bibitem{R18}
H. Tanaka \textit{et al.}, ``A precise on-chip voltage generator for a gigascale DRAM with a negative word-line scheme,'' \textit{IEEE Journal of Solid-State Circuits}, vol. 34, no. 8, pp. 1084--1090, Aug. 1999, doi: 10.1109/4.777106.

\bibitem{R19}
M. Pe\v{s}i\'{c} \textit{et al.}, ``Variability sources and reliability of 3D -- FeFETs,'' \textit{2021 IEEE International Reliability Physics Symposium (IRPS)}, Monterey, CA, USA, 2021, pp. 1--7, doi: 10.1109/IRPS46558.2021.9405118.

\bibitem{R20}
J.-M. Sallese and V. Meyer, ``The ferroelectric MOSFET: a self-consistent quasi-static model and its implications,'' \textit{IEEE Transactions on Electron Devices}, vol. 51, no. 12, pp. 2145--2153, Dec. 2004, doi: 10.1109/TED.2004.839113.

\bibitem{R21}
F.-X. Liang \textit{et al.}, ``A Physics-Based Model for Oxide--Semiconductor-Based Ferroelectric Field-Effect Transistors,'' \textit{IEEE Transactions on Electron Devices}, vol. 71, no. 7, pp. 4397--4402, July 2024, doi: 10.1109/TED.2024.3408776.

\bibitem{R22}
D. Chen \textit{et al.}, ``Antiferroelectric Phase Evolution in Hf$_x$Zr$_{1-x}$O$_2$ Thin Film Toward High Endurance of Non-Volatile Memory Devices,'' \textit{IEEE Electron Device Letters}, vol. 43, no. 12, pp. 2065--2068, Dec. 2022, doi: 10.1109/LED.2022.3217813.

\bibitem{R23}
S. Subhechha \textit{et al.}, ``Ultra-low Leakage IGZO-TFTs with Raised Source/Drain for $V_t > 0$ V and $I_{\mathrm{on}} > 30$ $\mu$A/$\mu$m,'' \textit{2022 IEEE Symposium on VLSI Technology and Circuits}, Honolulu, HI, USA, 2022, pp. 292--293, doi: 10.1109/VLSITechnologyandCir46769.2022.9830448.

\bibitem{R24}
G. Yeap \textit{et al.}, ``2nm Platform Technology Featuring Energy-Efficient Nanosheet Transistors and Interconnects Co-Optimized with 3DIC for AI, HPC and Mobile SoC Applications,'' \textit{2024 IEEE International Electron Devices Meeting (IEDM)}, San Francisco, CA, USA, 2024, pp. 1--4, doi: 10.1109/IEDM50854.2024.10873475.

\bibitem{R25}
N. Ramaswamy \textit{et al.}, ``NVDRAM: A 32Gb Dual Layer 3D Stacked Non-volatile Ferroelectric Memory with Near-DRAM Performance for Demanding AI Workloads,'' \textit{2023 International Electron Devices Meeting (IEDM)}, San Francisco, CA, USA, 2023, pp. 1--4, doi: 10.1109/IEDM45741.2023.10413848.

\end{thebibliography}
\end{document}